\documentstyle[12pt]{article}
\begin{document}
\begin{flushright}
{\large \bf IFUP-TH 34/94}
\end{flushright}
\vspace{0.5cm}
\begin{center}
\Large\bf  The Effective Non-Relativistic Hamiltonian\\
 for  Electron-Nucleus Interaction\\

\end{center}
\vspace{1.0cm}
\begin{center}
\large\bf R.Ya.Kezerashvili\footnote{Department of Physics, Tbilisi
State University, 380028, Tbilisi, Republic of Georgia.}
\end{center}
\vspace{0.5cm}
\begin{center}
Dipartimento di Fisica, Universita'di Pisa,\\
Piazza Torricelli 2, 56100 Pisa, Italy,\\
and INFN, Sezione di Pisa,\\
Piazza Torricelli 2, 56100 Pisa, Italy
\end{center}
\begin{abstract}
\bf We present the effective  Hamiltonian for an electron-nucleus
interaction in non-relativistic limit up to the second order in the inverse
nucleon mass. This Hamiltonian takes into account the distortion of
electron waves and allows to calculate its effect as well as matrix elements
for off-shell transitions in $(e,e'p)$ reactions.

\end{abstract}
\newpage
  The quasielastic $(e,e^{'}p)$ knockout reactions have been the object of
a large number of experimental and theoretical studies during the last 20
years.
The continuous improvements of the experimental apparatus and the quality of
the electron beams have allowed to collect  precise information on
single-particle properties of nuclei. Together with experimental efforts,
a precise theoretical treatment has developed, which correctly takes
into account all the main ingredients of this process [1].
The basic theoretical treatment  used
non-relativistic bound nucleon and continuous proton wave functions [1-7].
In particular, the single particle wave functions are generated by
Hartree-Fock
procedure and outgoing distorted proton wave function are taken as the
eigenfunction of phenomenological optical potential, obtained through a fit
to elastic proton-nucleus scattering. An alternative approach is to use a
non-relativistic random phase approximation model to calculate the
nuclear states and thereby avoid the use of an optical potential
[8,9]. On the other hand, in Refs. 10-13 a fully
relativistic approach has been adopted. Generally, in a relativistic
approach the bound state nucleon wave functions are the solutions of the Dirac
equation in the potential wells, derived  from a relativistic
mean-field Hartree calculation, since the continuous proton wave function is
derived using relativistic optical potentials calculated from a Dirac
phenomenological global fit to the elastic proton-nucleus scattering
observable. The diagram method was suggested in Ref. 14 and it allows
consequently to consider many particle short-range correlations in the
$(e,e^{'}p)$ reaction.

 A precise treatment of $(e,e^{'}p)$ reactions needs to include, besides the
outgoing proton distortion, also the electron distortion. The interaction of
the electrons with the field of the nuclei changes the incoming and outgoing
electron wave functions, and more realistic distorted-wave description must
be,  consequently, introduced in the $(e,e^{'}p)$ reaction. The problem is well
known and several efforts have been done to solve it. In Ref. 15 the
numerical solution of the Dirac equation for the electron in the nuclear
Coulomb field has been obtained with a
phase shift analysis based on a partial wave expansion. Assuming that the
virtual photon emitted by the electron is absorbed by a single nucleon, the
distorted wave functions of the electrons are obtained in ref. 12 by
numerically solving the Dirac equation in  presence of the static
Coulomb potential of the nuclear charge distribution.  It is necessary to
note that there is disagreement between [11] and [12], even if they use
the same general approach.

It is desirable to use an analytic approach to the Coulomb distortion of
the incoming and outgoing electron waves in the $(e,e^{'}p)$ reactions because
of the rather large and difficult numerical complications. Such approach have
been developed and applied in Refs. 4-7. It is based on a high energy
approximation for the electron wave function [16].  The eikonal
electron wave function is expanded  up to first  [4] and second order
[5,6] in $Z\alpha$  $(\alpha =1/137)$. The  cross section was calculated in
a non-relativistic approach, based on the impulse approximation.

The purpose of the present work is to construct an electron-nucleus
interaction Hamiltonian in the non-relativistic limit, taking into account
the distortion of the electron waves in the quasielastic electron
scattering on nuclei. This Hamiltonian allows  for carrying out an analytic
calculation of the electron waves distortion effect in the $(e,e^{'}p)$
reaction.

We make the following assumptions:

i. the incoming electron wave is distorted by the static Coulomb field of the
target nucleus.

ii. the resulting distorted electron wave interacts with a single nucleon
inside  the nucleus and knocks it out.

iii. the outgoing electron is distorted by the static Coulomb field of the
residual nucleus.

Thus, our treatment includes distortion of the electron waves due to the static
nuclear Coulomb field and we restrict ourself to one photon exchange. Such
correction is proportional to $Z/137$ and becomes critical for heavy
nuclei.  It implies  neglecting terms arising from the exchange of two or
more photons in the scattering event, terms which are supposed to give a small
contribution.

The covariant interaction between electrons and relativistic nucleons is,
of course, well known and the Dirac equation for a nucleon in an given
arbitrary external electromagnetic field is

\begin{equation}
  \Bigl[\,\gamma_{\mu}\partial_{\mu} + M -ieF_{1}A_{\mu}\gamma_{\mu}+
            \frac{e\,K\,F_{2}}{2\,M}\partial_{\nu}A_{\mu}\sigma_{\mu\nu}\,
       \Bigr]\Psi =0\,,
\end{equation}

where

\begin{equation}
 \sigma_{\mu\nu}=-\frac{i}{2}(\gamma_{\mu}\gamma_{\nu}-
\gamma_{\nu}\gamma_{\mu}).
 \end{equation}
In formula (1), $F_{1}$ and $F_{2}$ are the electromagnetic nucleon
form-factors, having usual normalization, $K$ is the anomalous magnetic
moment of the nucleon in nuclear magneton, $M$ is the nucleon mass and the
field strength are given through the four-vector potential
$A_{\mu}=(\mbox{\boldmath $A$},A_{4})\equiv(\mbox{\boldmath $A$},i\phi)$.

If we restrict to the non-relativistic limit, in which nucleons interact like
non-relativistic  Pauli particles, we need to reduce the Dirac equation
(1) to a form involving only two-component spinors for the nucleons. The
technique for carrying  out the non-relativistic limit of Dirac operators
involves  the construction of a series of the successive unitary
transformations, whose product is known as Foldy-Wouthuysen transformation
[17]. This transformation decouples equation (1) into two two-component
equations, one of which  reduces to the non-relativistic description, while the
other describes the negative energy states. By following McVoy and Van Hove
[18], applying the Foldy-Wouthuysen transformation to eq. (1) and retaining
terms up to  $M^{-2}$ order in the positive energy equation, finally the
Hamiltonian

  \[ H=-i\,e\,F_{1}\,A_{4}-\frac{e\,F_{1}}{2\,M}
 \bigl(\,\mbox{\boldmath $p\cdot A$}+
  \mbox{\boldmath $A$}\cdot \mbox{\boldmath $p$}\,\bigr)-\]

  \[\frac{e\,(F_{1}+K\,F_{2})}{2\,M}(\mbox{\boldmath $\sigma\cdot$} \mbox
   {\boldmath $H$})\,+\]

\[\frac{e\,(F_{1}+2\,K\,F_{2})}{8\,M^{2}}\Bigl(\mbox{\boldmath $\sigma
\cdot$} \bigl(
[\,\mbox{\boldmath $p$}\times \mbox{\boldmath $E$}\,]-[\mbox{\boldmath $E$}
\times \mbox{\boldmath $p$}\,]\,\bigr)\Bigr)\,-\]

\begin{equation}
  \frac{e\,(F_{1}+2\,K\,F_{2})}{8\,M^{2}} \,div\mbox{\boldmath $E$}
\end{equation}
is obtained. In the above expression $\mbox{\boldmath $p$}$ and
 $\mbox{\boldmath $\sigma$}$ are the  momentum and Pauli matrices for the
nucleon and the fields
$\mbox{\boldmath $E$}$ and $\mbox{\boldmath $H$}$ are generated by the
four-vector potential $A_{\mu}$

\begin{equation}\mbox{\boldmath $E$}=i\,grad\,A_{4}-\frac{\partial
\mbox{\boldmath $A$}}{\partial t}\,,\quad
         \mbox{\boldmath $H$}=curl\mbox{\boldmath $A$}\,.
\end{equation}
The Hamiltonian (3) describe the interaction of the non-relativistic nucleon
with an arbitrary electromagnetic field.

 In the plane wave approximation, when the incoming and the outgoing electron
 are described by  plane waves, $A_{\mu}$ will be the Moller potential [19].
By putting the Moller potential into equation (3) and (4), we obtain the
McVoy
and Van Hove Hamiltonian [18]. This Hamiltonian is well known and it has
been widely  used for investigations of the $(e,e^{'}p)$ processes.

What is the effect of the distortion, caused by the interaction of the target
and the residual nuclei  with the incoming and the outgoing electron waves?
Failing a complete calculation, a simple prescription is usually adopted to
include a first contribution  of electron distortion. The electron plane
wave is replaced by $e^{i\mbox{\boldmath $k^{'}r$}}$ [20] with

\begin{equation}
     \mbox{\boldmath $k^{'}$}=\mbox{\boldmath $k$}+ U(0)\mbox{\boldmath
$k$}/k
\,,
\end{equation}
where  $\mbox{\boldmath $k$}$ is the electron momentum and  $U(0)$ is a mean
value of the electromagnetic nuclear potential. In
Ref. 21 the electron plane wave was replaced by the plane wave with
effective momentum $\mbox{\boldmath $k^{'}$}$ and amplitude
$k^{'}/k$. This well known
result is called the effective momentum approximation. An high energy
electron wave function was obtained in Ref. 22 and the final expression for
the distorted electron wave is represented by a plane wave with the changed
phase and moduli. The corresponding expressions for this function is expanded
in powers of $Z\alpha$  and retaining terms up to
$Z\alpha$ and $(Z\alpha)^{2}$ are introduced in Refs.4-6 and they represent
the plane wave with the changed phase and amplitude too. So, in any case,
we can conclude, that from a
mathematical point of view, the distortion changes the amplitude and the
phase of the plane electron wave. This leads to the corresponding reduced
electron density and current. Thus, in  general , the electromagnetic
potential generated by the distorted electron density and current can be
written in the form

\[
  A_{4}=i\frac{4 \pi e\,u^{+}_{f}\,u_{i}}{q_{\mu}^{2}}\,V(\mbox{\boldmath $r$})
\,e^{iq_\mu r_\mu + i\Phi(\mbox{\boldmath $r$})}\,, \]

\begin{equation}
  \mbox{\boldmath $A$}= \frac{4\pi e\,u^{+}_{f}\, \mbox{\boldmath $\alpha$}\,
  u_{i}}{q_{\mu}^{2}}\,V(\mbox{\boldmath $r$})\,
e^{iq_\mu r_\mu + i\Phi(\mbox{\boldmath $r$})}\,,
\end{equation}
where $u$ is the Pauli  spinor for the free electron,
$\mbox{\boldmath $\alpha$}$ is the Dirac velocity matrix and $q_{\mu}$ is the
four-momentum transfer. In formulas (6) $V(\mbox{\boldmath $r$})$ and
$\Phi(\mbox{\boldmath $r$})$ are amplitude and
phase functions, reproducing the distortion of the incoming and  the outgoing
electron waves. Those functions have continuous second order derivatives.
Putting  $V(\mbox{\boldmath $r$})=1$ and $\Phi(\mbox{\boldmath $r$})=0$
into (6), we obtain the Moller potential.
Substituting the electromagnetic potential (6), induced by the distorted
electron waves, into eqs. (3) and (4),  we finally obtain

\[
H^{'}=\frac{4\,\pi\,e^{2}}{q_{\mu}^{2}}
  <\,u_{f}|\,F_{1}\,V(\mbox{\boldmath $r$})\,
e^{iq_\mu r_\mu + i\Phi(\mbox{\boldmath $r$})}\,- \]

\[
\frac{F_{1}}{2\,M}\Bigl(\,(\mbox{\boldmath $p\cdot \alpha$})\,
V(\mbox{\boldmath $r$})\,e^{iq_\mu r_\mu + i\Phi(\mbox{\boldmath $r$})}+
V(\mbox{\boldmath $r$})\,e^{iq_\mu r_\mu + i\Phi(\mbox{\boldmath $r$})}
(\mbox{\boldmath $p\cdot \alpha$})\Bigr)\,- \]

\[
\frac{F_{1}+K\,F_{2}}{2\,M}\,\Bigl\{ i\Bigl(\mbox{\boldmath $\sigma\cdot$} [\,
(\, \mbox{\boldmath $q$}+\bigtriangledown \Phi(\mbox{\boldmath $r$})\,)
\mbox{\boldmath $\times\alpha$}
\,]\,\Bigr)\,V(\mbox{\boldmath $r$})\,
e^{iq_\mu r_\mu + i\Phi(\mbox{\boldmath $r$})}\,+\]

\[
  \Bigl(\mbox{\boldmath $\sigma\cdot$} [\,\bigtriangledown
  V(\mbox{\boldmath $r$})\mbox{\boldmath $\times\alpha$}
\,]\,\Bigr)\,e^{iq_\mu r_\mu + i\Phi(\mbox{\boldmath $r$})}\,\Bigr\}\, +
 \]
\[
\frac{F_{1}+2KF_{2}}{8\,M^{2}}\,\Bigl\{\,i
  \mbox{\boldmath $\sigma\cdot$}\Bigl(
[\,\mbox{\boldmath $p\times$}
(\omega \mbox{\boldmath $\alpha$}-\mbox{\boldmath $q$}-\bigtriangledown
\Phi(\mbox{\boldmath $r$})
  )\,]\,
V(\mbox{\boldmath $r$})\,e^{iq_\mu r_\mu + i\Phi(\mbox{\boldmath $r$})}\,-
\]

\[
V(\mbox{\boldmath $r$})\,e^{iq_\mu r_\mu + i\Phi(\mbox{\boldmath $r$})}\,
[\,( \omega \mbox{\boldmath $\alpha$}-\mbox{\boldmath $q$}-\bigtriangledown
 \Phi(\mbox{\boldmath $r$}))\times \mbox{\boldmath $p$}]\,\Bigr)\,-
\]

\[
  \mbox{\boldmath $\sigma\cdot$}\Bigl(
[\,\mbox{\boldmath $p\times$} \bigtriangledown V(\mbox{\boldmath $r$})\,]
  e^{iq_\mu r_\mu + i\Phi(\mbox{\boldmath $r$})}  -
  e^{iq_\mu r_\mu + i\Phi(\mbox{\boldmath $r$})}
  [\,\bigtriangledown V(\mbox{\boldmath $r$})\mbox{\boldmath $\times p$}\,]\,
  \Bigr)\,\,\Bigr\}\,-
 \]

\[
\frac{F_{1}+2K\,F_{2}}{8\,M^{2}}\,\Bigl\{\,(\mbox{\boldmath $q$}+
 \bigtriangledown \Phi(\mbox{\boldmath $r$}))^{2}\,V(\mbox{\boldmath $r$})\,
e^{iq_\mu r_\mu + i\Phi(\mbox{\boldmath $r$})}\,-\]

\[
 2ie^{iq_\mu r_\mu + i\Phi(\mbox{\boldmath $r$})}
(\mbox{\boldmath $q$}+\bigtriangledown \Phi(\mbox{\boldmath $r$})\,)
\mbox{\boldmath $\cdot$}\bigtriangledown
  V(\mbox{\boldmath $r$})\,-
\]

\begin{equation}
iV(\mbox{\boldmath $r$})\,
e^{iq_\mu r_\mu + i\Phi(\mbox{\boldmath $r$})}
\bigtriangledown ^{2}
 \Phi(\mbox{\boldmath $r$})-
e^{iq_\mu r_\mu + i\Phi(\mbox{\boldmath $r$})}
\bigtriangledown ^{2}V(\mbox{\boldmath $r$})\,\Bigr\}\,|\,u_{i}>\,,
\end{equation}
where $\mbox{\boldmath $\alpha$}$ acts in the space of the electron spinors
and $\omega$ is the energy transfer.
Eq. (7) introduce the effective Hamiltonian of electron-nucleon interaction
in nuclear space, which describes the interaction of electron
waves distorted by nuclei  with  non-relativistic nucleons. When we go over
to the consideration of a nucleus, the result of eq. (7) must be summed
over the A nucleons.

The first three terms of the Hamiltonian (7), which describe the Coulomb,
convection current and spin or magnetization current interactions with
distorted electron wave are of the order of unity and $M^{-1}$. These terms
provide the dominant contributions, especially at lower momentum transfers.
The two other terms, of order $M^{-2}$, do contribute since the momentum
transfers we deal with may be quite large. They are important at large
momentum transfers. If we substitute $V(\mbox{\boldmath $r$})=1$ and
$\Phi(\mbox{\boldmath $r$})=0$
into equation (7) the McVoy and Van Hove Hamiltonian [18] is obtained.

The effective momentum transfer $\mbox{\boldmath $q$}+\bigtriangledown \Phi(
\mbox{\boldmath $r$})$ in eq.(7) differs from $\mbox{\boldmath $q$}$ both in
magnitude and in direction. When the change of the direction is neglected,
the elastic scattering condition is obtained.
If we choose $V(\mbox{\boldmath $r$})=1$ and
$\Phi(\mbox{\boldmath $r$})=U(0)(\mbox{\boldmath $\hat{q}\cdot r$})$, where
$U(0)=3Z\alpha/2R$ with $R=(5/3)^{1/2}<r>^{1/2}$ for the nucleus with charge
$Z$ and rms-radius $<r>^{1/2}$, we are considering the distorted electron waves
in the approximation used in Ref. 20. In this case, eq. (7) simplifies, because
terms
containing $\bigtriangledown V(\mbox{\boldmath $r$})$ disappear. The
same simplification will happen  in the high energy approximation
[18], but taking the amplitude $k{'}/k$ for the plane wave. The comparison the
of equations (6) with the electromagnetic potential obtained in Ref. 4,
allows to find corresponding expressions for $V(\mbox{\boldmath $r$})$ and
$\Phi(\mbox{\boldmath $r$})$. Using those expressions in the Hamiltonian
(7), the distortion of the electron waves will be taken into account, based
on the approximation of Ref. 16.

In Ref. 3  the non-relativistic Hamiltonian for free electron-nucleon
scattering up to the fourth order in the inverse nucleon mass was obtained.
Starting from this Hamiltonian and using the same procedure, we obtain the
effective
non-relativistic Hamiltonian,  describing the interaction of the electron
wave distorted by nuclei  with the nucleon, corrected up to  $M^{-4}$.

Using the Hamiltonian (7) for the description of $(e,e{'}p)$ reactions, the
off-shell transition matrix elements can be calculated  and the
distortion of electron wave in the initial and  final states can be included.
\vspace{1cm}

I wish to thank Prof. A.Fabrocini for helpful discussion and the
Department of Physics of Pisa University for the kind hospitality.

\newpage
{\bf References:}

\begin{enumerate}

\item S Boffi, C.Giusti and F.D.Pacati. Phys.Rep. {\bf 226} (1993) 1.
\item S Boffi, C.Giusti and F.D.Pacati. Nucl.Phys. {\bf A 386} (1982) 599.
\item C.Giusti and F.D.Pacati. Nucl.Phys. {\bf A 336} (1980) 427.
\item C.Giusti and F.D.Pacati. Nucl.Phys. {\bf A 476} (1987) 717.
\item C.Giusti and F.D.Pacati. Nucl.Phys. {\bf A 485} (1988) 461.
\item M.Traini, S.Turck-Chi\`{e}ze and A.Zghiche. Phys.Rev. {\bf C 38} (1988)
2799.
\item M.Traini. Phys.Lett. {\bf B 213} (1988) 1.
\item J.Ryckebusch, K.Heyde, D.Van Neck and M.waroquier. Phys.Lett. {\bf B
216} (1989) 252.
\item M.Cavinato, M.Marangoni, and A.M.Saruis. Z.Phys. {\bf A 335} (1990)
401.
\item A.Picklesimer, J.W.Van Orden and S.J.Wallace. Phys.Rev. {\bf C 32}
(1985) 1312.
\item J.P.McDermott. Phys.Rev.Lett. {\bf 65} (1990) 1991.
\item Yanhe Jin, D.S.Onley and L.E.Wright. Phys.Rev. {\bf C 45} (1992)
1311.
\item Yanhe Jin, J.K.Zhang, D.S.Onley and L.E.Wright. Phys.Rev. {\bf C 47}
(1993) 2024.
\item R.I.Jibuti and R.Ya.Kezerashvili Yad.Fiz. {\bf 22} (1975) 975
[Sov.J.Nucl.
Phys. {\bf 22} (1975) 508.
\item R.D.Viollier and K.Alder. Helv.Phys.Acta. {\bf 44} (1971) 77.
\item J.Knoll. Nucl.Phys. {\bf A 223} (1974) 462.
\item L.L.Foldy and S.A.Wouthuysen. Phys.Rev. {\bf 78} (1950) 29.
\item K.W.McVoy and L.Van Hove. Phys.Rev. {\bf 125} (1962) 1034.
\item C.Moller. Z. Phys. {\bf 70} (1931) 786.
\item W.Czy\^{z} and K.Gottfried. Ann. of Phys. {\bf 21} (1963) 47.
\item R.Rosenfelder. Ann. of Phys. {\bf 128} (1980) 188.
\item F.Lenz and R.Rosenfelder. Nucl.Phys. {\bf A 176} (1971) 513.

\end{enumerate}
\end{document}